\documentclass[12pt]{iopart}

\usepackage{graphicx}
\usepackage[dvipdfmx,colorlinks=true,linkcolor=blue,citecolor=blue,urlcolor=blue]{hyperref}
\newcommand{\singlet}{{}^1\mathrm{S}_0}
\newcommand{\triplet}{{}^3\mathrm{P}_2}
\newcommand{\Er}{E_R^{\mathrm{Yb}}}
\newcommand{\cm}{\mathrm{cm}}
\newcommand{\micron}{\mu\mathrm{m}}
\newcommand{\nm}{\mathrm{nm}}
\newcommand{\msec}{\mathrm{ms}}
\newcommand{\s}{\mathrm{s}}
\newcommand{\eqref}[1]{equation\,(\ref{#1})}

\begin{document}

\title[Collisional stability of localized Yb($\triplet$) atoms immersed in a Fermi sea of Li]{Collisional stability of localized Yb(\boldmath{$\triplet$}) atoms immersed in a Fermi sea of Li}

\author{Hideki Konishi, Florian Sch\"{a}fer, Shinya Ueda and Yoshiro Takahashi}
\address{Department of Physics, Graduate School of Science, Kyoto University, Kyoto 606-8502, Japan}
\ead{h.konishi@scphys.kyoto-u.ac.jp}
\date{\today}

\begin{abstract}
We establish an experimental method for a detailed investigation of inelastic collisional properties between ytterbium (Yb) in the metastable $\triplet$ state and ground state lithium (Li).
By combining an optical lattice and a direct excitation to the $\triplet$ state
we achieve high selectivity on the collisional partners.
Using this method we determine inelastic loss coefficients in collisions between $^{174}$Yb($\triplet$) with magnetic sublevels of $m_J=0$ and $-2$ and ground state $^6$Li to be
$(4.4\pm0.3)\times10^{-11}~\cm^3/\s$ and $(4.7\pm0.8)\times10^{-11}~\cm^3/\s$, respectively.
Absence of spin changing processes in Yb($\triplet$)-Li inelastic collisions at low magnetic fields is confirmed by inelastic loss measurements on the $m_J=0$ state.
We also demonstrate that our method allows us to look into loss processes in few-body systems separately.
\end{abstract}

\vspace{2pc}

\maketitle

\section{Introduction}
Impurities play crucial roles in condensed-matter physics such as Anderson localization~\cite{Anderson1958}, the Kondo effect~\cite{Kondo1964}, and Anderson's orthogonality catastrophe~\cite{Anderson1967}.
A better understanding of these phenomena through experiments still remains an important task.
Ultracold atomic gases in optical lattices~\cite{Bloch2008,Giorgini2008} can provide outstanding opportunities to study impurity problems with excellent controllability,
where the impurities are introduced in controlled ways:
by optical fields such as incommensurate optical lattices~\cite{Fallani2007,Roati2008} and optical speckles~\cite{Billy2008,Kondov2011},
or by atomic impurities~\cite{Gunter2006,Ospelkaus2006,Gadway2010}.

An ultracold mixture of ytterbium (Yb) and lithium (Li) is one of the promising systems to realize an atomic impurity system.
Loaded in a suitable optical lattice, Yb atoms are deeply localized in lattice sites
while Li atoms remain itinerant over the whole system because of their extreme mass imbalance $m_{\mathrm{Yb}}/m_{\mathrm{Li}}\approx 29$.
An experimental challenge to study impurity problems with Yb-Li mixtures lies in the control of their collisional properties.
It is theoretically predicted that
Feshbach resonances (FRs) between Yb and Li in their respective ground states are too narrow to precisely tune the inter-species interaction~\cite{Brue2012}.
On the other hand, the metastable excited $\triplet$ state of Yb offers an interesting possibility to control interactions between Yb and Li.
FRs between ground and excited $\triplet$ state Yb atoms have been observed in several isotopes recently~\cite{Kato2013,Taie2016},
demonstrating the feasibility of working with FRs between different orbitals.
In addition to usual mechanisms of FRs as in alkali atoms~\cite{Chin2010},
the observed resonances arise from anisotropy effects in their interactions~\cite{Kotochigova2014}.
In consideration of these recent results,
it is reasonable to also expect some useful FRs in the Yb($\triplet$)-Li system.

Indeed, several theoretical investigations of FRs in $^{174}$Yb($\triplet$)-$^6$Li~\cite{Gonzalez-Martinez2013} and $^{171}$Yb($\triplet$)-$^6$Li systems~\cite{Chen2015} are reported.
The existence of several FRs is predicted.
Considering the complexity of the calculations involved, together with uncertainty of the constructed inter-atomic potentials,
experimental feedback is indispensable to refine quantitative predictions of resonance positions and inelastic loss rates.
On the experimental side, a mixture of $^{174}$Yb($\triplet$, $m_J=-1$) and $^6$Li was realized at a few~$\mu\mathrm{K}$~\cite{Khramov2014},
and variations of the inelastic loss rate of the $^{174}$Yb($\triplet$, $m_J=-1$)-$^6$Li collisions for $100\textrm{--}520\,\mathrm{G}$ were observed~\cite{Dowd2015}.
Further experimental investigations of inelastic collisions at a low temperature regime at other magnetic fields and with other magnetic substates should give further insight into Yb($\triplet$)-Li collisions.
Especially, FRs in a $^{174}$Yb($\triplet$, $m_J=-2$)-$^6$Li system are theoretically predicted~\cite{Petrov2015} based on optimized potentials obtained from the $m_J=-1$ experimental results.
Experimental confirmation of those predictions are as of yet lacking.

In this paper, we realize a system of localized Yb atoms with controllable internal states immersed in a Fermi sea of Li in a three-dimensional optical lattice.
We investigate inelastic losses in detail in a quantum degenerate Yb-Li mixture.
Instead of using spin-polarized fermionic isotopes immersed in a Fermi degenerate gas of Li as in a $^{40}$K-$^6$Li mixture~\cite{Spiegelhalder2009},
performing the experiments in a deep optical lattice
and using direct excitation from the Yb ground to the excited $\triplet$ state,
we achieve high selectivity on the collisional partners and high flexibility in the target state preparation.
This method allows us to work also with bosonic Yb isotopes.
Accordingly the results presented in this work provide a general survey on inelastic collisional properties of Li with one or several Yb atoms in ground and $\triplet$ states.

\section{Sample preparation and detection}
A quantum degenerate mixture of $^{174}$Yb and $^6$Li is prepared as described in~\cite{Hara2011,Hara2014}.
Typically we obtain a mixture of a Bose Einstein condensation of $8\times 10^4$ Yb atoms and a Fermi degenerate gas of $2.5\times10^4$ Li atoms
after evaporative cooling in a crossed optical far-off-resonance trap (FORT).
The Fermi gas of Li consists of the two spin states in the ground $F=1/2$ state.
The Li temperature is $T_{\mathrm{Li}}=500\,\mathrm{nK}$
and $T_{\mathrm{Li}}/T_{\mathrm{F}}\simeq 0.4$,
where $T_{\mathrm{F}}$ is the Fermi temperature.
The trap frequencies of Yb and Li are
$2\pi \times(70, 90, 153)\,\mathrm{Hz}$ and $2\pi \times(519,852,1440)\,\mathrm{Hz}$, respectively.

At the final stage of evaporative cooling, the Yb cloud sits about $6.5\,\micron$ below the Li cloud due to their different gravitational sag, which results in a reduced spatial overlap between them.
To compensate for this, we apply an intensity gradient of a laser whose wavelength is $532\,\nm$.
This is complementary to using a magnetic field gradient as in~\cite{Hansen2013}.
The laser field acts as an attractive potential for Yb and a repulsive one for Li.
This gravitational sag compensation beam (GCB) has a waist of $75\,\micron$ and is pointing about $38\,\micron$ above the atomic cloud.
The Yb cloud is pulled up by about $3\,\micron$ at a GCB power of $900\,\mathrm{mW}$.
Due to its strong confinement within the FORT, Li is nearly unaffected and does not move.
We linearly ramp up the GCB in $100\,\msec$ followed by $100\,\msec$ holding at the end of evaporation to prevent heating and oscillations of the atomic sample.

\begin{figure}[bt]
\centering
	\includegraphics{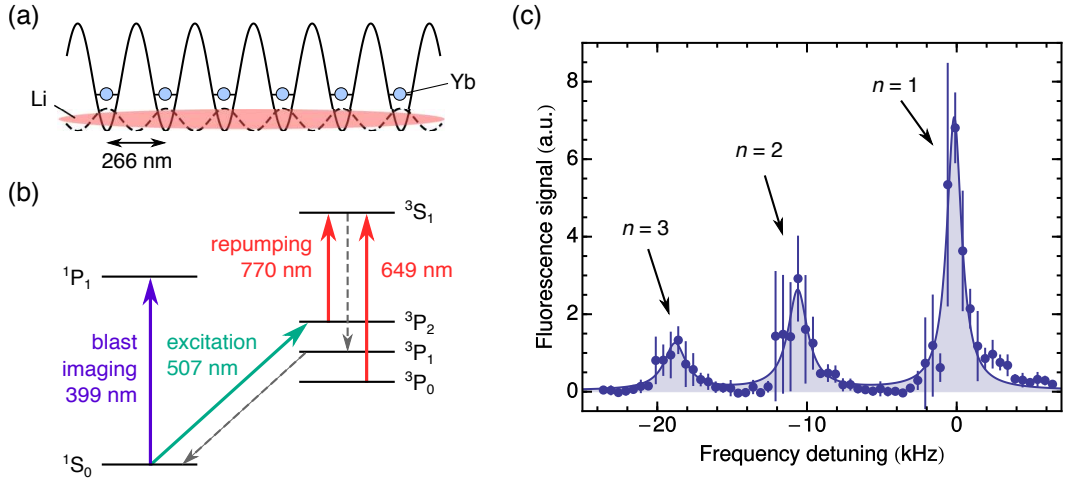}
\caption{(a) Delocalized Li and localized Yb atoms in an optical lattice.
Even though lattice sites alternate for Yb and Li (solid and dashed lines, respectively)
because of the opposite signs of their polarizabilities at $\lambda_\mathrm{L}=532\,\nm$,
the Li atoms are delocalized enough to overlap with the localized Yb atoms due to the shallow specific potential.
(b) Relevant energy levels of $^{174}$Yb for $\triplet$ state preparation and detection.
(c) A typical $\singlet\to\triplet(m_J=0)$ excitation spectrum of $^{174}$Yb with $^6$Li in an optical lattice at $15\,\Er$.
Resonances of single, double, and triple occupancies ($n=1, 2, 3$) in the lattice are separated due to interatomic interaction.
The solid line denotes the fits with Lorentzian functions for each of the resonances.}
\label{spectroscopy}
\end{figure}

We then adiabatically load a Yb-Li quantum degenerate mixture into a 3D optical lattice with wavelength $\lambda_{\mathrm{L}} = 532\,\nm$ and form a Yb Mott insulator.
The optical lattice is adiabatically ramped up to $15\,\Er$ in $200\,\msec$,
where $E_R=\hbar^2(2\pi/\lambda_{\mathrm{L}})^2/(2m)$ is the recoil energy with atomic mass $m$.
The ratio of $s=V_{\mathrm{L}}/E_R$, the lattice depth divided by the recoil energy, for Yb and Li is $s_{\mathrm{Yb}}/s_{\mathrm{Li}}=21.4$.
At $s_{\mathrm{Yb}}=15$ we have $s_{\mathrm{Li}}=0.7$ at which the Bloch state is well delocalized in the system.
This ensures, even though at $\lambda_\mathrm{L}=532\,\nm$ their polarizabilities have opposite signs, reasonable overlap between the delocalized Li and the localized Yb atoms (see \fref{spectroscopy}(a)).

Here we describe $\triplet$ state preparation and detection methods.
See \fref{spectroscopy}(b) for the relevant energy levels.
A portion of the ground state Yb atoms is directly excited to the $\triplet$ state by a $0.1\textrm{--}1\,\msec$ laser pulse at a resonant wavelength of $507\,\nm$.
The excitation laser has a linewidth of about $100\,\mathrm{Hz}$.
An applied magnetic bias field lifts the degeneracy of the $m_J$ sublevels
and allows for $m_J$ selective excitation to the $\triplet$ state.
Typically we apply $282\,\mathrm{mG}$ for the excitation of the $m_J=0$ state and $200\,\mathrm{mG}$ for the $m_J=-2$ state.
To reduce inhomogeneous broadening it is important to choose the field orientation for each magnetic sublevel
because the polarizabilities of different $m_J$ states have different dependence on the angle between the direction of the magnetic field and the polarization of the laser field~\cite{Hara2014}.
For further reduction of inhomogeneous broadening the GCB is ramped down to $0\,\mathrm{mW}$ while the lattice goes from 10 to $15\,\Er$ in $50\,\msec$.
Since Yb atoms are already pinned at lattice sites at $10\,\Er$,
the spatial overlap between Yb and Li remains restored even with the GCB turned off at this point.

For the detection of the $\triplet$ atoms,
we first remove the ground state Yb atoms from the trap
by a $0.5\textrm{--}1\,\msec$ laser pulse resonant to the $\singlet\to{}^1\mathrm{P}_1$ transition.
The atoms in the $\triplet$ state are repumped to the ground state via the $^3$S$_1$ state
by simultaneous applications of two laser pulses resonant
to the $\triplet\to{}^3\mathrm{S}_1$ and $^3\mathrm{P}_0\to{}^3\mathrm{S}_1$ transitions with a duration of $1\,\msec$ (cf.~\fref{spectroscopy}(b)).
Finally, the atoms returned to the ground state are recaptured by a magneto-optical trap (MOT) operating on the strong $\singlet\to{}^1\mathrm{P}_1$ transition.
The fluorescence intensity from the MOT is detected and is proportional to the number of repumped atoms.
\Fref{spectroscopy}(c) shows a typical $\singlet\to\triplet(m_J=0)$ excitation spectrum of the Yb Mott insulator at $15\,\Er$ immersed in a Fermi degenerate gas of Li.
The formation of a Mott insulator state is reflected by the spectrum with well-resolved resonance peaks shifted by on-site interactions~\cite{Kato2016}.
We note that the corresponding spectrum taken without Li (not shown) is basically identical,
in agreement with previous observations~\cite{Hara2014}.
In the measurements below, we selectively excite Yb atoms in singly, doubly, and triply occupied ($n=1,\,2,\,3$) sites by properly setting the excitation laser frequency.

\section{Results}
\subsection{Yb($\triplet$)-Li inelastic collisions}
We measure the loss of Yb($\triplet$, $m_J=0$) atoms by the collisions with Li.
The experimental procedure is as follows;
after the lattice depth reaching $15\,\Er$
the $507\,\nm$ excitation pulse resonant to Yb singly occupied ($n=1$) lattice sites is applied.
Remaining ground state Yb atoms are removed by $\singlet\to{}^1\mathrm{P}_1$ resonant light as described above.
This procedure allows us to exclude unwanted Yb($\triplet$)-Yb($\triplet$) and Yb($\triplet$)-Yb($\singlet$) collisions that have shown to have large inelastic loss rates on the order of $10^{-11}$ and $10^{-12}\,\cm^3/\s$, respectively~\cite{Uetake2012}.
It is also important to note that the the number of excited atoms is less than 10\% of that of Li
so that the Li density can be considered constant during the interaction time.
After a variable holding time of the Yb($\triplet$)-Li mixture in the lattice
we detect the Yb atoms remaining in the $\triplet$ state.
Li atoms are detected by absorption imaging at the same time.
For comparison, we do the same measurement for a sample without Li,
where Li atoms are removed from the trap by applying a laser pulse resonant to the Li D2 line with a duration of $1\,\msec$ before loading the lattice.

\begin{figure}[bt]
\centering
	\includegraphics{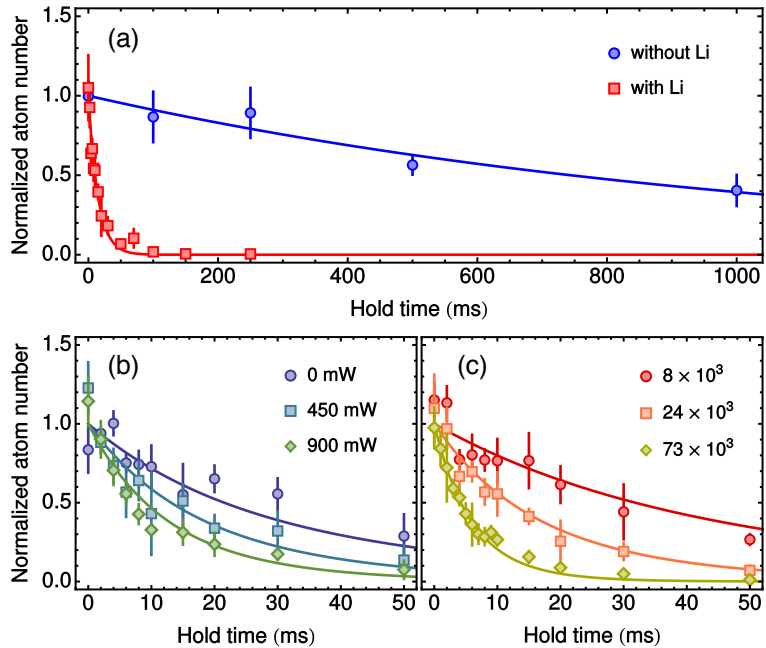}
\caption{(a) Time evolutions of Yb($\triplet$, $m_J=0$) atoms with and without Li.
The GCB power and the total number of Li are $900\,\mathrm{mW}$ and $2.5\times10^4$, respectively.
A stronger decay in the presence of Li is clearly observed.
Fits by exponential functions (solid lines) give decay time constants of
$\approx15\,\msec$ and $\approx1000\,\msec$, respectively.
Inelastic decays of Yb($\triplet$) atoms (b) at various GCB powers, and (c) with various total numbers of Li.
The decays become faster as Li density increases.
Solid lines denote fits to the data (see text).
Error bars indicate the standard deviation of three independent measurements.}
\label{mJ0}
\end{figure}

The result is shown in \fref{mJ0}(a).
A fast decay of Yb($\triplet$) atoms by the collision with Li is clearly observed.
Fits to the data with exponential functions give decay time constants of $\approx15\,\msec$ and $\approx1000\,\msec$
for the cases with and without Li, respectively.
To determine the Yb($\triplet$, $m_J=0$)-Li inelastic loss coefficient,
we repeat the measurement with various Li densities
by changing either GCB intensities or total numbers of Li.
We vary the GCB power between $0\sim900\,\mathrm{mW}$
and the total Li atom number between $8\sim73\times10^3$
by changing the initial Li loading time.
Results with representative GCB powers are shown in \fref{mJ0}(b).
The atoms decay faster as GCB power increases.
This also indicates the effectiveness of our GCB approach in which it lifts the Yb cloud up to the denser region of the Li cloud.
\Fref{mJ0}(c) shows results with representative total numbers of Li atoms.
Faster decay with more Li atoms is clearly discerned.

\begin{figure}[tb]
\centering
	\includegraphics{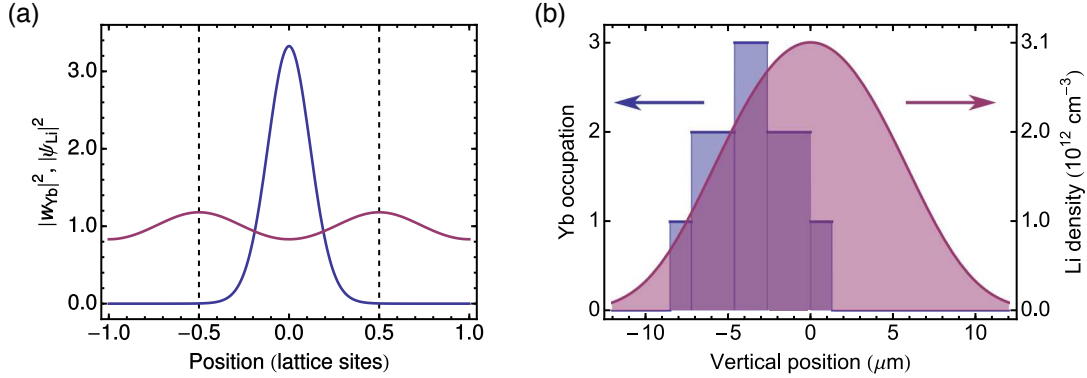}
\caption{(a) The Wannier state of Yb (blue) and the Bloch state of Li (red) at $s_\mathrm{Yb} (s_\mathrm{Li})=15 (0.7)$.
The density correction factor $\xi$ is determined by the overlap integral of them in a single lattice site (between -0.5 and 0.5 lattice sites).
(b) Mott shell structure of the Yb ground state (blue) and density distribution of Li (red) along the vertical direction.
Examples with $N_{\mathrm{Yb}}=1.0\times10^5$ and $N_{\mathrm{Li}}=2.5\times10^4$ at GCB power $900\,\mathrm{mW}$ are displayed.
The Yb($\triplet$) atoms can be assumed to be equally distributed in the $n=1$ Mott shell volume,
because we excite the atoms in $n=1$ shell selectively.}
\label{density}
\end{figure}

We fit the datasets and determine the inelastic loss coefficient in the following way.
In the absence of Yb($\triplet$)-Yb($\triplet$) and Yb($\triplet$)-Yb($\singlet$) collisions thanks to the optical lattice and the occupancy selective excitation scheme,
the decay of Yb($\triplet$) atoms is described by the Yb($\triplet$)-Li inelastic decay term as a dominant process;
\begin{equation}
	\dot{n}_{\mathrm{Yb}}=-\alpha n_{\mathrm{Yb}}-\beta \xi n_{\mathrm{Li}}n_{\mathrm{Yb}},
\label{ndot}
\end{equation}
where $n_{\mathrm{Yb}}$ and $n_{\mathrm{Li}}$ are the density of Yb($\triplet$) and Li, respectively,
$\alpha$ is the one-body loss rate, and $\beta$ is the Yb($\triplet$)-Li inelastic loss coefficient.
We have further introduced a Li density correction factor $\xi$.
It accounts for the reduced density of the Li Bloch wave function at Yb sites for a lattice depth of
$s_\mathrm{Li}=0.7$ (see \fref{density}(a)).
The correction factor $\xi$ is determined by the three dimensional overlap integral of the Wannier state of Yb and the Bloch state of Li in a single lattice site;
\begin{equation}
	\xi=\int_{-d/2}^{d/2} |w_\mathrm{Yb}(r)|^2 |\psi_{\mathrm{Li}}(r)|^2\,\mathrm{d}^3r.
\label{overlap}
\end{equation}
Here $w_\mathrm{Yb}$ is the Wannier state of Yb,
$\psi_\mathrm{Li}$ is the Bloch state of Li,
and $d=266\,\nm$ is the lattice spacing.
Evaluating the Yb Wannier function we include that the lattice depth for Yb($\triplet$) in each direction is slightly different due to the dependence of its polarizability on the angle between the magnetic field orientation and the laser polarization.
For the $m_J=0$ state the correction factor is calculated to be $\xi=0.66$.

Considering the number of Yb($\triplet$) is less than 10\% of that of Li atoms,
we regard $n_{\mathrm{Li}}$ as time independent as mentioned above.
Therefore, the time evolution of the number of Yb($\triplet$) is expressed as
\begin{equation}
	N_{\mathrm{Yb}}(t)=\int n_{\mathrm{Yb}}(r, t=0)e^{-(\alpha+\beta \xi n_{\mathrm{Li}}(r))t}\,\mathrm{d}^3r.
\label{fit}
\end{equation}

\Fref{density}(b) shows a typical Mott shell structure of the Yb ground state and a density distribution of Li along vertical direction with GCB power $900\,\mathrm{mW}$.
Since we selectively excite Yb atoms in the $n=1$ Mott shell,
we assume that the Yb($\triplet$) atoms are equally distributed in $n=1$ shell volume.
The one-body loss rate is determined to be $1/\alpha=(900\pm250)\,\msec$ from the fit to the data without Li.
We evaluate $\beta$ using a bootstrap method.
All datasets are fitted 100 times by \eqref{fit} with $\beta$ being a common parameter among them.
Each time the Li cloud size, the numbers of Yb and Li, and  the Yb vertical position are randomly chosen in the ranges of $\pm10\%$ (roughly corresponds to the Li temperature $\pm150\,\mathrm{nK}$), $\pm10\%$, and $\pm0.5\,\micron$, respectively.
The mean and the standard deviation of the fit results yield
$\beta=(4.4\pm0.3)\times10^{-11}\,\cm^3/\s$ for the $m_J=0$ state.
Solid lines in \fref{mJ0}(b) and (c) show the fit results.

\begin{figure}[tb]
\centering
	\includegraphics{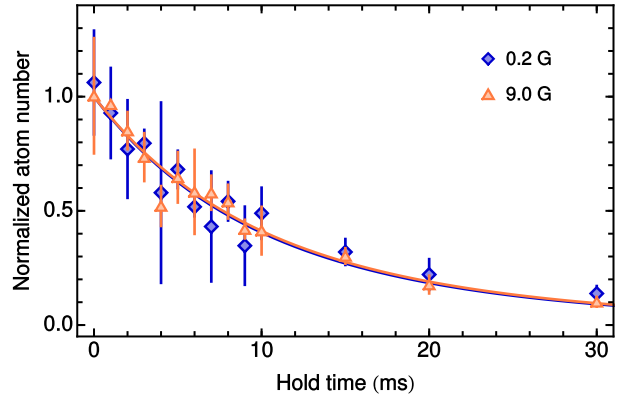}
\caption{Inelastic decays of Yb($\triplet$, $m_J=-2$) with Li at $200\,\mathrm{mG}$ and $9.0\,\mathrm{G}$.
Solid lines are fits to the data with \eqref{fit}.
Both decay curves are rather identical.
This does not support the previous theoretical prediction of a FR at around $10\,\mathrm{G}$.}
\label{mJ-2}
\end{figure}

For the determination of the inelastic loss coefficient for the $m_J=-2$ state,
possible contributions from both $n=1$ and $2$ shells are included.
They arise from excitation uncertainties caused by small intra-species interaction in combination with a significant sensitivity to magnetic field noise of this state.
Our error budget accounts for this by also allowing partial to full excitation in the $n=2$ shell in the bootstrap analysis.
The analysis with $\xi=0.64$ results in $\beta=(4.7\pm0.8)\times10^{-11}\,\cm^3/\s$.
This value is two orders of magnitude smaller than that predicted in~\cite{Petrov2015}
and better matches the prediction in~\cite{Gonzalez-Martinez2013}.
In the former report, the authors also predict an increase of $\beta$ with a $^{174}$Yb($\triplet$, $m_J=-2$)-$^6$Li FR at around $10\,\mathrm{G}$,
while the latter predicts a decrease of $\beta$ for $0\textrm{--}50\,\mathrm{G}$.
To check these, we compare the decays of Yb($\triplet$, $m_J=-2$) at magnetic fields of $200\,\mathrm{mG}$ and $9.0\,\mathrm{G}$ (\fref{mJ-2}).
The magnetic field is swept to the desired value in $1\,\msec$ after the excitation.
The experimentally obtained decay curves at both magnetic fields are almost identical,
highlighting the continuing challenges in a theoretical treatment of the problem.
Our result provides additional input to refine the required inter-atomic potentials.

\begin{figure}[bt]
\centering
	\includegraphics{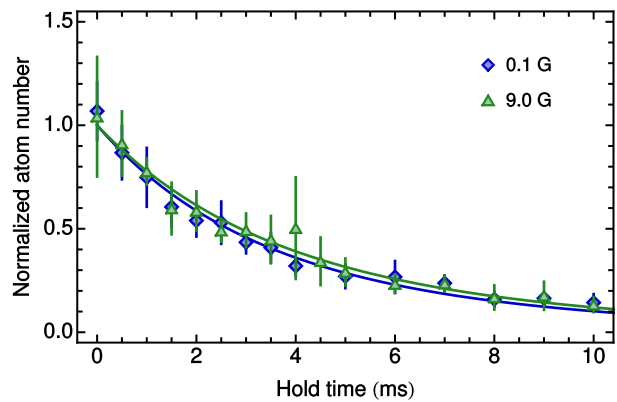}
\caption{Inelastic decays of Yb($\triplet$, $m_J=0$) with Li at $100\,\mathrm{mG}$ and $9.0\,\mathrm{G}$.
Solid lines are fits to the data with \eqref{fit}.
No differences are observed below and above the threshold magnetic field $500\,\mathrm{mG}$.
This excludes spin changing collision dominance in the Yb($\triplet$)-Li inelastic collisions.}
\label{mJ0B}
\end{figure}

To give further insight into the Yb($\triplet$)-Li inelastic collisions, we investigate the inelastic decay channels of the $\triplet$ atoms by the collision with Li.
Possible decay processes are spin changing, fine structure changing, and principal quantum number changing collisions.
Considering energy and momentum conservation in Yb($\triplet$)-Li inelastic collisions,
a decayed Yb atom carries away only $m_{\mathrm{Li}}/(m_{\mathrm{Li}}+m_{\mathrm{Yb}})\approx 3$\% of the released energy.
If spin changing collisions dominantly occur,
there is a magnetic field threshold beyond which $\triplet(m_J>-2)$ atoms in the process $m_J \to -2$ gain more energy (3\% of the magnetic field dependent Zeeman splitting) than the lattice and FORT support.
As a result, we would expect faster decays of $m_J>-2$ states at higher magnetic fields.
Since the Zeeman splitting of the $\triplet$ state of $^{174}$Yb is
$k_{\mathrm{B}}\times100\,\mu\mathrm{K/G}$ per $\Delta m_J=1$,
the threshold is about $500\,\mathrm{mG}$ for $m_J=0$ at a lattice depth of
$15\,\Er=k_{\mathrm{B}}\times2.9\,\mu\mathrm{K}$.
We compare the Yb($\triplet$, $m_J=0$) decays at $100\,\mathrm{mG}$ and $9.0\,\mathrm{G}$ (\fref{mJ0B}).
We do not find any significant differences between the two cases.
Therefore, we conclude that the decay of Yb($\triplet$) by inelastic collisions with Li at low magnetic field is dominated by fine structure changing or principal quantum number changing collisions.

\subsection{Site-occupancy selective loss measurements}
Besides the Yb($\triplet$)-Li inelastic collision measurements,
we demonstrate that our method using an optical lattice and a direct excitation allows us to study collisional processes site-occupancy selectively.
First, we measure the decay of $\triplet$ state in doubly occupied ($n=2$) sites.
By selectively exciting Yb atoms in $n=2$ sites to the $\triplet$ state,
Yb($\triplet$)-Yb($\singlet$) collisions become detectable
while Yb($\triplet$)-Yb($\triplet$) and higher order collisions are inhibited.
We use the $m_J=-2$ state that is stable against the collision with Yb($\singlet$)~\cite{Uetake2012}.
The experimental procedure is as in the above measurements apart from the absence of the Yb($\singlet$) blast pulse before having the holding time.

\begin{figure}[tb]
\centering
	\includegraphics{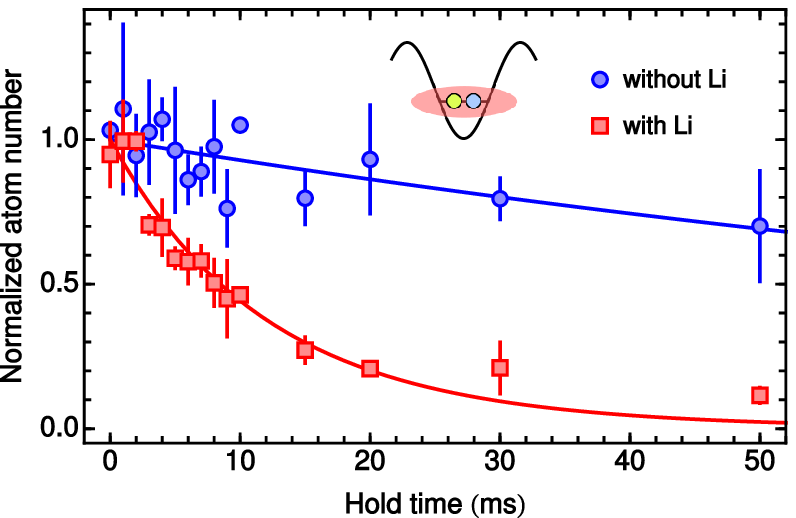}
\caption{Time evolutions of Yb($\triplet$, $m_J=-2$) in $n=2$ sites.
The solid lines denote fits to the datawith an exponential function for the data without Li
and with \eqref{fit} for the data with Li.
Any particular behavior in a Yb($\triplet$)-Yb($\singlet$) system with Li is not confirmed.
Inset: Sketch of the situation.
In a single lattice site Yb($\singlet$) (blue), Yb($\triplet$) (yellow), and delocalized Li (red) interact.}
\label{ybybli}
\end{figure}

The result is shown in \fref{ybybli} together with the case with Li for comparison.
The decay model is described by \eqref{fit} with $\alpha$ modified by the collision with Yb($\singlet$) atoms.
From an exponential fit to the data without Li,
$\alpha$ is determined to be $1/\alpha=(135\pm20)\,\msec$ in agreement with the previous result in~\cite{Uetake2012}.
The fit to the data with Li by \eqref{fit} yields
$\beta=(5.4\pm1.0)\times10^{-11}\,\cm^3/\s$,
a value similar to the one obtained above for Yb($\triplet$, $m_J=-2$)-Li collisions in absence of Yb($\singlet$) atoms.
This demonstrates that the Li-induced inelastic decay of the two-atom state of Yb($\singlet$)$+$Yb($\triplet$) can be approximated reasonably by assuming Li to only affect the Yb($\triplet$) atoms.

\begin{figure}
\centering
	\includegraphics{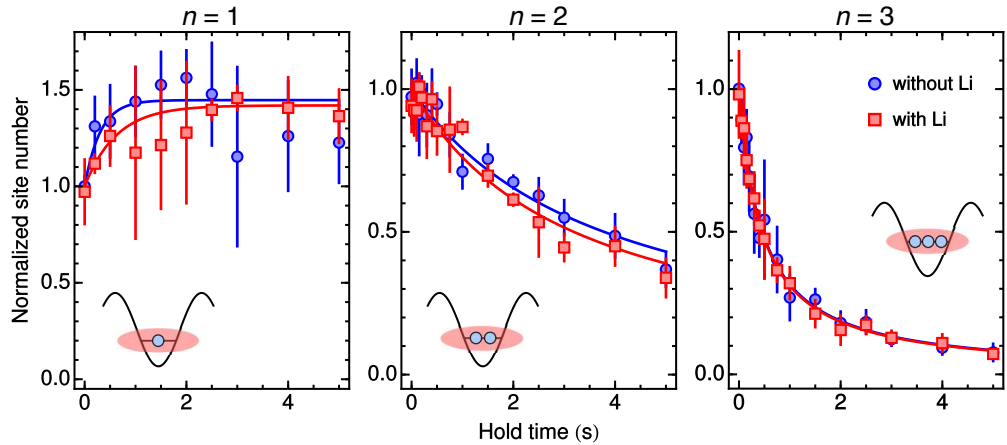}
\caption{Time evolutions of the number of singly, doubly, and triply occupied sites measured by site-occupancy selective excitation.
The number of $n=1$ sites increases by about $50\%$ in a few seconds.
Decays of $n=2$ and $3$ sites are observed.
These behaviors should be related not only to three-body decay processes but also to complication of tunneling and on-site interactions.
In any occupation order, no notable differences between the cases of presence and absence of Li are observed.
Solid lines are guides to the eye.}
\label{nyb}
\end{figure}

To complete the picture, we also investigate by selective excitation the time evolutions of only ground state Yb atoms in singly, doubly, and triply occupied lattice sites separately.
We have a variable hold time at $15\,\Er$ and then measure for each occupation number the remaining number of sites by site-occupancy selective excitation to the $m_J=0$ state.
Here we note that since our excitation method excites only one of the atoms in each lattice site,
not the number of atoms but the number of sites is measured.

The results are shown in \fref{nyb}.
An increase of the number of singly occupied sites
and decays of those of doubly and triply occupied sites are observed.
The observed behaviors should be attributed to an intricate dynamics where tunneling and interaction interplay as well as three-body decays.
In the case of singly and doubly occupied sites the dynamics is likely to be dominated by redistribution of the site occupation by hopping and due to heating effects induced by the lattice beams.
In triply occupied sites molecule formation also becomes possible.
In contrast to the systems including Yb($\triplet$) atoms, notable differences between the cases with and without Li are not observed in any occupation order.
This proves that in the Yb ground state the intra-species collisional properties are not significantly altered by the presence of Li.

\section{Conclusions and outlook}
We develop an experimental method combining a deep optical lattice and a direct excitation
to the $\triplet$ state to investigate Yb($\triplet$)-Li inelastic collisional properties in detail.
The $^{174}$Yb($\triplet$)-$^6$Li inelastic loss coefficients for $m_J=0$ and $-2$ states
are determined to be $(4.4\pm0.3)\times10^{-11}\,\cm^3/\s$
and $(4.7\pm0.8)\times10^{-11}\,\cm^3/\s$, respectively.
The obtained inelastic loss rate of $m_J=-2$ and its magnetic field dependence should provide stimulus to further improve current calculations on the Yb($\triplet$)-Li FR landscape.
Observed magnetic field independence of the inelastic loss rate with $m_J=0$ proves little contribution of spin changing processes to the decay of Yb($\triplet$) in collisions with Li.
Our method also allows us to investigate decays of atoms in one- or few-body systems separately.
In fact, we measure the time evolution of Yb($\triplet$) in $n=2$ sites,
and we find that the Li-induced inelastic decay of the two-atom state of Yb($\singlet$)$+$Yb($\triplet$) is well understood by the Li atoms affecting solely the Yb($\triplet$) state.
Further we successfully observe time evolutions of ground state Yb atoms in $n=1,\,2,\,\textrm{and}\,3$ sites separately
and confirm absence of the effect of Li on the intra-species collisional properties.

The experimental method presented in this work can serve as a tool in the search for Yb($\triplet$)-Li FRs by measuring variations of inter-species inelastic loss rates over a wide range of magnetic fields.
Also it is applicable to other isotopes.
Especially, fermionic Yb isotopes ($^{171}$Yb and $^{173}$Yb) are interesting candidates to search for inter-species FRs,
where both usual and anisotropy-induced FRs are expected to exist because of their hyperfine structures in the $\triplet$ states.
We plan to also make use of high-resolution spectroscopy on the $\singlet\to\triplet$ transition to measure scattering lengths between Yb($\triplet$) and Li at eventually confirmed FRs as performed in~\cite{Kato2013,Taie2016}.

\ack
We acknowledge K.\,Ono and A.\,Kell for experimental assistance.
This work was supported by the Grant-in-Aid for Scientific Research of JSPS
No.\,25220711, No.\,26247064, No.\,16H00990, and No.\,16H01053
and the Impulsing Paradigm Change through Disruptive Technologies (ImPACT) program by the Cabinet Office, Government of Japan.
H.\,K.\ acknowledges support from JSPS.

\section*{References}
\bibliography{library}

\end{document}